\documentclass[%
 reprint,
superscriptaddress,
 amsmath,amssymb,
 aps,
showkeys ]{revtex4-1}

\usepackage{graphicx}
\usepackage{dcolumn}
\usepackage{bm}

\newcommand{\CNRSAddress}{Univ. Grenoble Alpes, CNRS, Institut NEEL, F-38000 Grenoble, France}
\newcommand{\CEAAddress}{Univ. Grenoble Alpes, CEA, INAC, F-38000 Grenoble, France}

\begin{document}

\preprint{APS/123-QED}

\title{Nanowire growth and sublimation: CdTe quantum dots in ZnTe nanowires}

\author{M. Orr\`{u}}
\affiliation{\CNRSAddress} \affiliation{\CEAAddress}

\author{E. Robin}
\affiliation{\CEAAddress}

\author{M. Den Hertog}
\affiliation{\CNRSAddress}

\author{K. Moratis}
\affiliation{\CNRSAddress}

\author{Y. Genuist}
\affiliation{\CNRSAddress}

\author{R. Andr\'e}
\affiliation{\CNRSAddress}

\author{D. Ferrand}
\affiliation{\CNRSAddress}

\author{J. Cibert} \email{joel.cibert@neel.cnrs.fr} \affiliation{\CNRSAddress}

\author{E. Bellet-Amalric}
\affiliation{\CEAAddress}

\date{\today}

\begin{abstract}
The role of the sublimation of the compound and of the evaporation
of the constituents from the gold nanoparticle during the growth of
semiconductor nanowires is exemplified with CdTe-ZnTe
heterostructures. Operating close to the upper temperature limit
strongly reduces the amount of Cd present in the gold nanoparticle
and the density of adatoms on the nanowire sidewalls. As a result,
the growth rate is small and strongly temperature dependent, but a
good control of the growth conditions allows the incorporation of
quantum dots in nanowires with sharp interfaces and adjustable
shape, and it minimizes the radial growth and the subsequent
formation of additional CdTe clusters on the nanowire sidewalls, as
confirmed by photoluminescence. Uncapped CdTe segments dissolve into
the gold nanoparticle when interrupting the flux, giving rise to a
bulb-like (pendant-droplet) shape attributed to the Kirkendall
effect.

\end{abstract}

\maketitle

\section{\label{Intro}Introduction}

Fabricating semiconductor quantum dots (QD) embedded in a nanowire
(NW) constitutes a more flexible process than the usual
Stranski-Krastanow growth mode: it permits to combine various
materials with various strain configurations, and adjustable shape
and size. One interest of adjusting the aspect ratio and the strain
configuration \cite{Niquet,Zielinski} is that it allows one to shape
the hole state ("heavy hole" or "light hole"), without involving
complex structures\cite{Huo}. This has a strong impact on the
optical selection rules: emission diagram for classical photonic
devices, photovoltaics, and quantum devices for optical manipulation
of the electron or hole state in a doped quantum dot - these aspects
are particularly attractive for III-V and II-VI material
combinations. It also governs the spin properties.

The growth of NWs by molecular beam epitaxy (MBE) is generally
achieved thanks to a nano-sized catalyst, most often a gold droplet
(the so-called vapor-liquid-solid growth mode, or VLS): the elements
constituting the NW are trapped in the droplet where the nucleation
to form the semiconductor is more efficient than in the neighboring
vapor phase. The QD is inserted by switching the impinging flux for
a certain time, which controls the QD height, while the diameter of
the contact area between the catalyst and the NW controls the
diameter of the QD. For a good control of the QD shape, sharp
interfaces are needed, meaning that the problem of "reservoir
effect" (the progressive change of composition of the droplet when
switching the flux) has to be circumvented. Various solutions were
proposed. Silicon-germanium axial heterostructures were fabricated
using a solid catalyst instead of a liquid one (vapor-solid-solid
growth, or VSS, instead of VLS): the solubility is expected to be
lower \cite{Wen} and a model was proposed which assumes that surface
and interface diffusion takes place instead of diffusion through the
catalyst nanoparticle \cite{Cui}. In III-V heterostructures, the
reservoir effect is small, a few \%, when dealing with column-V
atoms \cite{Priante2015, Johansson2017} but the residual pressure in
the MBE chamber has to be mastered. The reservoir effect is more
severe when dealing with column-III atoms, and elaborate sequences
were designed to make it as small as possible \cite{Dick2012,
Priante2016}.

We address the growth of QDs embedded in NWs made of II-VI
semiconductors. Several aspects make such structures particularly
attractive. Light-hole emission was reported in such CdTe QDs in
ZnTe NWs \cite{Jeannin}. Single photon emission was observed up to
room temperature in CdSe QDs in ZnSe NWs \cite{Bounouar}. II-VI QDs
in NWs can incorporate a dilute magnetic
semiconductor\cite{Szymura}: in that case one can expect the
formation of a magnetic polaron around a hole trapped in the QD,
which constitutes a model system for ultimate magnetic objects where
the superparamagnetic state is induced by one or a small number of
carriers, so that it can be controlled by a low applied bias. Again,
the anisotropy is governed by the hole state - a feature reminiscent
of carrier induced ferromagnetism in a quantum well. Finally, CdTe
NWs have been used recently \cite{Kessel} as a template for the
growth of HgTe NWs in the quest for 1D topological insulators.

It has been recognized for some time that the growth window for CdTe
NWs, using MBE with a gold catalyst, is extremely narrow, with a
high temperature limit attributed to the sublimation of CdTe
\cite{Wojnar2017}. Indeed a dramatic drop of the growth rate is
observed when the temperature is increased by a few degrees above
$350^\circ$C. Our goal here is to make this observation more
quantitative and to unravel the role played in the growth of CdTe by
the sublimation of CdTe and the evaporation of Cd from the gold
catalyst, and the consequences for the properties of CdTe QDs.

\section{\label{Methods}Experimental methods and samples}

\subsection{\label{MBE}Growth by molecular beam epitaxy}

The details of the conditions used for the sample preparation and
the growth of NWs have been reported previously\cite{Rueda growth}.
The growth was achieved by MBE in a Riber 32 chamber, using solid
catalyst particles formed by dewetting a fraction of a monolayer of
gold deposited on a ZnTe buffer on a GaAs substrate. All fluxes were
calibrated prior to growth by measuring the RHEED oscillations on a
ZnTe or CdTe (001) test substrate: for instance the Zn flux (and the
Te flux) from the ZnTe cell is measured by the growth rate of a ZnTe
layer, in ML s$^{-1}$, with an additional Zn flux from a Zn cell
\cite{Arnoult}. The sample holder was aligned on the horizontal line
containing the CdTe cell (angle of incidence $\alpha$ in the
horizontal plane, $\tan \alpha=0.21$) while our ZnTe cell is on the
other cell line ($\tan \alpha=0.48$).

\subsection{\label{EDX-TEM}EDX and TEM}

We used energy dispersive x-ray (EDX) spectrometry coupled to a FEI
Tecnai Osiris scanning transmission electron microscope (TEM)
equipped with four Silicon Drift Detectors and operated at 200 kV.
The NWs were removed mechanically from the as-grown samples and
deposited on a holey carbon-coated copper grid. The EDX signal is an
hypermap where each pixel corresponds to the x-ray emission of atoms
along the electron beam path. We used the QUANTAX-800 software from
Bruker for background correction and deconvolution to extract the
contributions of the L lines of Te, Cd, Au, and K lines of Zn and O.
The absorption correction - for the typical size of the NWs - was
estimated to be negligible for Te, Zn, Cd and less than 10\% for O.
The cross-sections for each element are deduced from the so-called
$\zeta$-factors directly measured on our equipment at the same
operating conditions using reference samples of known composition
and thickness \cite{Lopez-Haro}.

High angle annular dark field (HAADF) high resolution scanning TEM
images were realized on a probe corrected FEI Titan Themis operated
at 200 kV.

\subsection{\label{Samples}Samples}

In this paper we discuss three samples grown by MBE on a ZnTe buffer
layer, in the VSS mode. They incorporate a series of CdTe-based
insertions, see Fig.~\ref{fig1}a, in a micrometer-long ZnTe NW:
sample I to test the CdTe-ZnTe interfaces, sample T to test the
effect of growth temperature, and sample R to measure the growth
rate by testing the effect of the duration of the CdTe pulse.

The three samples are:
\begin{itemize}
\item Sample I (interfaces) contains a series of thin CdTe insertions which were previously used as markers to
follow the growth of the ZnTe NW \cite{Rueda growth}; the intended
structure of this sample is shown schematically at the top of
Fig.~\ref{fig1}a.
\item Sample T (temperature) involves a series of CdTe pulses of identical duration, but the temperature was decreased by steps of $5^\circ$C at
the middle of each ZnTe sequence, down to $350^\circ$C for the last
insertion.
\item Sample R (rate) was designed to measure the growth rate by testing the effect of the duration of the CdTe pulse:
the duration was increased by a factor 2 for each insertion; the
nominal growth temperature was $375^\circ$C.
\end{itemize}
The last sequence was ZnTe for sample I and CdTe for sample T and R.
The position of the insertions along the NW is such that the growth
of the CdTe insertion and of the surrounding shells is essentially
due to the flux impinging onto the catalyst and onto the sidewalls
of the NW, with no contribution of adatoms diffusing from the
substrate \cite{Rueda growth}.   In all cases we observe both
zinc-blende NWs, with a cone shape, and wurtzite NWs, with a
cylinder shape \cite{Rueda2014}. The local composition has been
determined by a quantitative modeling \cite{Rueda2016} of EDX
spectrometry, and by the geometrical phase analysis (GPA)
\cite{hytchGPA,rouviereGPA} of TEM images.

Section \ref{Results} describes the experimental results obtained on
these three samples. Based on these results, section  \ref{Model}
proposes a modified version of the model previously used to describe
the growth of ZnTe NWs \cite{Rueda growth}: we incorporate the
sublimation of CdTe and the evaporation of Cd from the gold
nanoparticle. Finally section  \ref{Discussion} discusses the
consequences for the axial and radial growth rates and the influence
on the luminescence properties of a CdTe dot in ZnTe NW; it also
describes the impact on the nanoparticle, which acquires a bulb-like
shape which we attribute to the Kirkendall effect induced by a
redissolution of CdTe.

\section{\label{Results}Experimental results}

\begin{figure*}
 \centering
 \includegraphics[width=2\columnwidth]{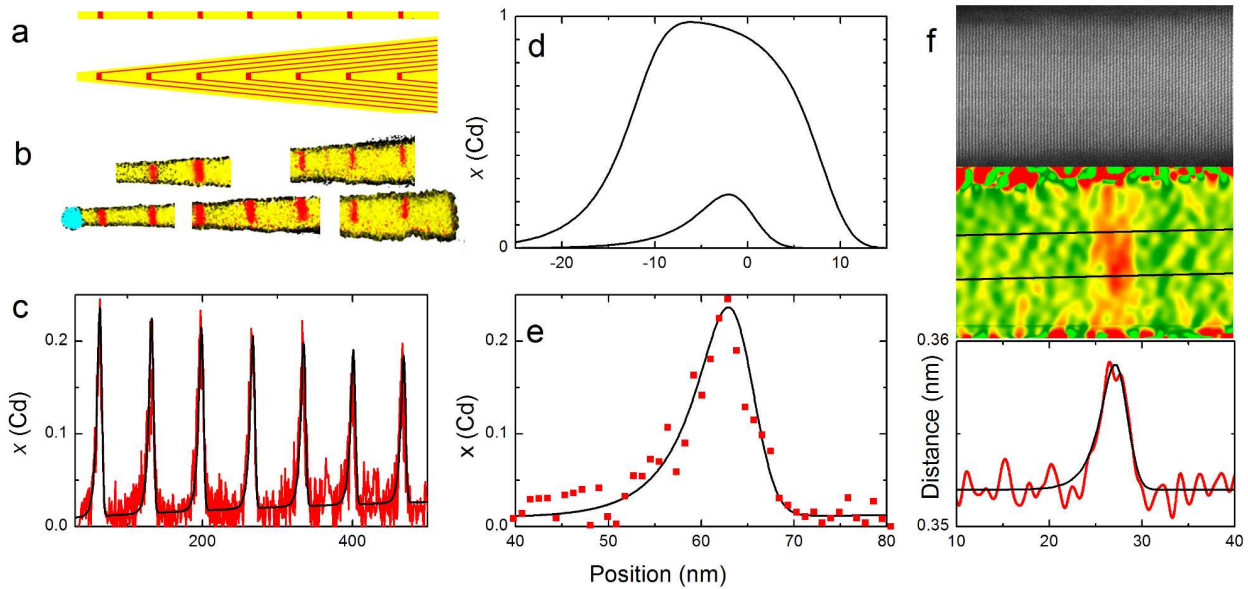}
\caption{Sample I: (a) scheme of the structure of the NW without
(top) and with (bottom) radial growth; (b) EDX map along one NW,
showing the area where O (black), Au (cyan), Cd (red) and Zn
(yellow) exceed an arbitrary concentration; (c) EDX axial profile of
Cd in the same NW, with fit; (d) two examples of the fitting profile
obtained using Eq.~\ref{Eq interface}, one with an insertion length
as in (e) and another one with a 10 times longer insertion; (e) EDX
profile of the topmost insertion in (c), and fit with Eq.~\ref{Eq
interface}, $d=2$~nm, $\tau=4$~nm , $\sigma=2$~nm; (f) TEM image
(top) and GPA profile (bottom) averaged on the band indicated in the
image); the fit uses $d=1.4$~nm, $\tau=1.5$~nm , $\sigma=1$~nm. }
\label{fig1}
\end{figure*}

Figure~\ref{fig1} displays the result of the EDX study of one NW
from sample I. The profile (Fig.~\ref{fig1}e) is not symmetrical and
features a tail towards the NW tip; the Cd content $x$ of
Cd$_x$Zn$_{1-x}$Te at position $z$ can be fitted with two interfaces
separated by a distance $d$, $x(z)=x_i(z-d/2)-x_i(z+d/2)$, each
interface being broadened by a Gaussian of width $\sigma$ and an
exponential of length $\tau$ extending along the growth direction
towards the tip. While a more sophisticated model of the interface
could be elaborated \cite{Priante2015, Priante2016}, as suggested by
the different growth rates of CdTe and ZnTe and the different values
of their formation energy, it would require an accurate knowledge of
the thermodynamics of Cd, Zn and Te in the gold nanoparticle and
this is beyond the scope of the present study. Going on with the
simple hypothethis of an exponential tail, the convoluted profile is
obtained by a straightforward calculation as

\begin{eqnarray}\label{Eq interface}
x_i(z)=&& \frac{1}{2}\left[1-\textrm{erf}
(\frac{z}{\sqrt{2}\sigma})\right] \nonumber\\
&&+\frac{1}{2}\left[1+\textrm{erf}
(\frac{z-\sigma^2/\tau}{\sqrt{2}\sigma})\right]\exp(-\frac{z-\sigma^2/2\tau}{\tau})
\end{eqnarray}

Figure~\ref{fig1}d shows two examples of such a profile, the one
which provides a good fit to the measured EDX profile
(Fig.~\ref{fig1}e), and the other one with a ten-times thicker
insertion, which allows a better identification of the role of the
exponential tails.

Finally, Fig.~\ref{fig1}c and \ref{fig1}e show that the same
individual profile, with the same parameters, well describes the
overall axial profile of the NW, provided we take into account the
presence of the radial ZnTe-CdTe multishell structure, as
schematized in the bottom part of Fig.~\ref{fig1}a. This multishell
structure is due to lateral growth: for instance, during the growth
of a CdTe insertion, Cd adatoms which are present on the NW
sidewalls and do not reach the gold nanoparticle may induce some
lateral growth of CdTe and form a CdTe shell over the previously
grown sections of the NW; of course this shell will be absent from
the sections grown subsequently. In a symmetrical way, a ZnTe shell
is formed during the growth of each ZnTe section. The resulting
structure, with an increasing number of individual shells from the
NW tip to its base, is schematically displayed at the bottom of
Fig.~\ref{fig1}a, where the lateral growth has been exaggerated. The
EDX analysis provides the average Cd content (with respect to Cd+Zn)
averaged over the electron beam path. The presence of Zn shells
creates the tapering and this additional Zn content along the
electron beam path decreases the signal measured at each insertion
in Fig.~\ref{fig1}c; the fit assumes a factor of 1.6 between the
diameter along the electron beam path at the basis and at the tip of
the NW. The presence of Cd shells slightly contributes to the
tapering, and creates the rising background between the insertions
in Fig.~\ref{fig1}c: the fit assumes that a CdTe shell of thickness
$\sim$~0.2~nm (half a monolayer) is added by the growth of each
insertion.

This NW from sample I had been transferred onto a grid for the EDX
study and it may have been broken during the harvesting process.
Other NWs were observed by TEM on a cleaved edge \cite{Rueda growth}
so that we are sure that they grew perpendicular to the substrate,
and that they are complete although their basis is hidden by the 2D
layer grown at the same time as the NWs. Profiles were obtained by
GPA: GPA measures the crystal plane distance, also averaged over the
electron beam path, with respect to its value measured in the ZnTe
sequence. An example is displayed in Fig.~\ref{fig1}f. A good fit of
the experimental profile is obtained with slightly lower values of
the parameters, $d$ and $\tau\approx1.5$~nm and $\sigma\approx1$~nm.
The larger values of the parameters measured by EDX are probably
partly of instrumental origin (GPA requires a perfect alignment of
the NW in the TEM, EDX does not); they may be due also to a slight
angle of the NW axis with respect to the substrate normal during the
growth: notice the rather strong tapering, and the asymmetric shell
surrounding the CdTe insertions. We conclude that this kind of
profile is validated by EDX and GPA, and it will be used in the
following.

\begin{figure}
 \centering
 \includegraphics[width=1\columnwidth]{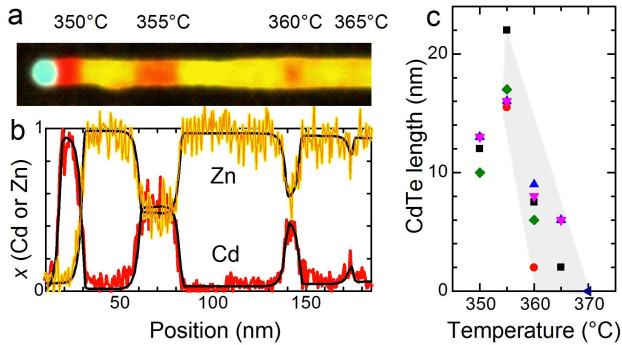}
\caption{Sample T: (a) EDX map (cyan: Au; yellow: Zn; red: Cd) and
TEM image of a NW (scale bar 20 nm); (b) axial profile of Cd and Zn
concentrations in the same NW, with fit (black solid lines) using
the same interface profile $\tau=2.0$~nm, $\sigma=0.5$~nm and
different values of length $d$; (c) length $d$ of the CdTe
insertions measured on several NWs, as a function of growth
temperature; the grey area suggests the trend followed by the capped
insertions, while the uncapped ones are shorter.} \label{fig2}
\end{figure}

Figure~\ref{fig2} displays the results of the EDX study of a NW from
sample T, with CdTe insertions grown at different temperatures: TEM
image, EDX map, axial EDX profiles for Cd and Zn (total=1), and its
fit with Eq.~\ref{Eq interface}, $\tau=2.0$~nm , $\sigma=0.5$~nm and
different values of $d$. The CdTe pulse was the same for all
insertions, about 4 times larger than in sample I. It is clear that
the CdTe dot grown at $355^\circ$C is much longer than the dot at
$360^\circ$, while the dot at $365^\circ$ almost disappeared. This
trend was confirmed on 6 other NWs from sample T, with either
wurtzite or zinc-blende structure, with the results gathered in
Fig.~\ref{fig2}c. Two additional NWs in the series are long enough
to confirm that dots grown at $370^\circ$C are beyond the detection
limit by EDX.

\begin{figure}
 \centering
 \includegraphics[width=1.0\columnwidth]{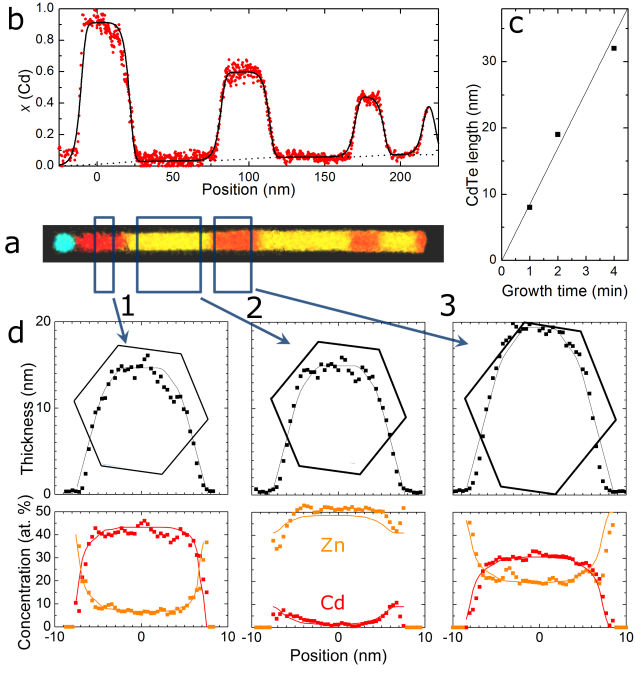}
\caption{Sample R: (a) EDX map (cyan: Au; yellow: Zn; red: Cd);
(b)axial profile of Cd and fit (solid line) using the baseline shown
by the dotted lines and elemental contributions using the same
interface profile $\tau=3.0$~nm, $\sigma=2.0$~nm, and different
values of length $d$; (c) interface distance $d$ as a function of
the CdTe pulse length; the uncapped insertion is omitted; (d) radial
profiles (total, Cd in red, Zn in orange) across the three
rectangles shown in (a), and fit using the hexagonal cross-sections
as indicated (the inner line indicates the boundary between the core
and the multishell area; the internal boundaries of the multishell
are not shown). } \label{fig3}
\end{figure}

Figure~\ref{fig3} displays the results obtained on sample R, where
the length of the CdTe pulse was progressively increased. The EDX
map (Fig.~\ref{fig3}a) indeed suggests that the length of the CdTe
insertion correspondingly increases, but for the last, uncapped
insertion. The axial profile (Fig.~\ref{fig3}b) and the plot in
Fig.~\ref{fig3}c confirm the length-duration dependance.

The radial profiles (Fig.~\ref{fig3}d) reveal interesting features.
Summing all contributions, we can plot the local thickness of the
NW, which is best fitted assuming an hexagonal cross section. In the
analysis\cite{Rueda2016}, the fitting parameters are the overall
orientation of the facets with respect to the electron beam (the
vertical axis in the schemes of Fig.~\ref{fig3}d, which was not
oriented with respect to the crystal structure in the NW) and the
distance between opposite facets. Profile d(2) reveals a ZnTe core
with a shell enriched in CdTe; the thickness of the shell is
difficult to ascertain, but the total amount of CdTe corresponds to
less than a monolayer of pure CdTe. Profile d(1) points to an almost
pure CdTe core with a thin shell enriched in ZnTe. Profile d(3)
demonstrates an increase of the thickness by lateral overgrowth
(tapering), with a ZnTe-rich shell; the apparent Cd content of the
core is reduced, part of it is due to the presence of the thicker
ZnTe shell, but we cannot exclude some radial diffusion of Cd from
the core to the shell. Note that all inter-facet distances increase
from d(1) to d(2) and d(2) to d(3), but not identically, so that the
shape of the hexagons is changed ( $\pm$10\% with respect to the
regular hexagon). Although we have no definitive interpretation, we
have already noted \cite{Rueda growth} that the shape of the
nanoparticle is affected by random changes during the growth. This
affects both radial and axial growth (and it probably accounts for a
part of the length fluctuations observed in Fig.~\ref{fig2}c and
Fig.~\ref{fig3}c). It does not invalidate the general conclusion of
the present study.

The temperature dependance of the CdTe growth rate, as demonstrated
in Fig.~\ref{fig2}c, is completely different from the weak
temperature dependance that we found for ZnTe \cite{Rueda growth}.
In the following we show that this difference is due for a part to
the cell-NW geometry, and for the greatest part to the evaporation
of Cd from the nanoparticle and the sublimation of CdTe from the
sidewalls, which are both significant in this temperature range when
compared to the impinging flux.

\section{\label{Model}Model}

Elaborating on the model previously used to describe the growth of
ZnTe NWs \cite{Rueda growth}, we consider a NW with a quasi-sphere
nanoparticle of diameter $D_{NP}$. The NW diameter at the tip, noted
$D_{NW}$, may be different: in the case of ZnTe, our previous
analysis of the growth rate \cite{Rueda growth} lead us to assume
$D_{NW} /D_{NP}\approx0.6$. In the case of CdTe, we will keep the
same values as a reasonable starting point. The normal component of
the incident flux, $J_s$, has been measured from RHEED (reflexion
high energy electron diffraction) oscillations during growth on a
test substrate; it is expressed as the growth rate of a 2D layer, in
nm~s$^{-1}$. The NW is normal to the substrate and the angle of
incidence of the flux is $\alpha$: in our MBE chamber, $\tan
\alpha$=0.21 for the CdTe cell and 0.48 for the ZnTe cell. The NW
length is significantly larger than the diffusion length along the
sidewalls ($\lambda\approx80$~nm at $350^\circ$C for ZnTe
\cite{Rueda growth}), so that the diffusion of adatoms from the
substrate is negligible. Hence there are two contributions to the NW
growth: (1) the direct flux to the nanoparticle, $J_s \pi D_{NP}^2
/4 \cos \alpha$ (the quasi-full sphere intercepts the total flux
$J_s / \cos \alpha$ and not only its normal component $J_s $), and
(2) the flux to the NW sidewalls, $J_s \tan \alpha $, integrated
over the NW diameter and the diffusion length $\lambda$, $J_s \tan
\alpha ~D_{NW} \lambda $. Their contribution to the growth rate of
the NW of cross section area $ \pi D_{NW}^2 /4$ is thus

\begin{equation}\label{Eq NW}
V_{NP}^{in}=\frac{J_s }{\cos \alpha}\frac{D_{NP}^2}{D_{NW}^2},~
V_{SW}^{in}=\frac{J_s \tan \alpha}{\pi}\frac{~ 4 \lambda }{ D_{NW}},
\end{equation}

respectively. We neglect the contribution of the flux reflected or
re-emitted from the substrate \cite{Ramdani}.

We now estimate the effect of the two negative contributions,
evaporation from the nanoparticle and sublimation of CdTe adatoms
from the sidewalls, to finally propose a global balance.

Cadmium is a volatile species: the vapor pressure above Cd at
$350^\circ$C is $p_{Cd}^0=$40~Pa (compared to 2~Pa for Zn and 0.4~Pa
for Te$_2$ ). The flux evaporating from a unit surface area is
\cite{Glas2013} $k_ep_{Cd}~/~\sqrt{2\pi M k_B T}$ with $M$ the mass
of a Cd atom (112~g for $N_A$ atoms, with $N_A$ the Avogadro
constant), $k_B$ the Boltzmann constant, $T$ the temperature, and
$k_e$ a factor representing barrier effects (in principle smaller
than 1 but often omitted). In the case of a quasi-full sphere, this
flux is integrated over the surface area, $\pi D_{NP}^2$, and it
reduces the growth rate of the NW of cross section area $ \pi
D_{NW}^2 /4$ by a volume $a_0^3/4$ per Cd atom, where $a_0$=0.648~nm
is the CdTe lattice constant. The final reduction of the growth rate
is thus $V_{NP}^{out}=\frac{k_ep_{Cd}}{\sqrt{2\pi M k_B T}}a_0^3~
\frac{D_{NP}^2}{D_{NW}^2}$. With $p_{Cd}^0$=40~Pa at $350^\circ$C,
$\frac{p_{Cd}^0}{\sqrt{2\pi M k_B T}}a_0^3\approx0.3$~mm~s$^{-1}$
(millimeter per second). This illustrates the benefit of a gold
catalyst, which will reduce the Cd vapor pressure to
$p_{Cd}=a~p_{Cd}^0$ with an activity $a$ much smaller than the Cd
concentration $c$, and small enough to limit the evaporation of Cd
to reasonable values: low values indeed were reported \cite{Filby,
Bartha} for the activity of Cd in gold at low concentration (8 to
10\%), as low as 10$^{-5}$ for 8\% Cd in gold wires \cite{Bartha}.
With such values, the loss of material due to the evaporation is
reasonable, but not negligible. The temperature and concentration
dependance is discussed in the supplementary material \cite{Sup}: a
reasonable extrapolation in the frame of the regular solution model
is

\begin{eqnarray}\label{Eq pCd}
p_{Cd}(c,T)&&=3.3\times10^{10}~c\nonumber\\
&&\times\exp\left[-\left(12100+7800\frac{(1-c)^2}{1+0.27c}\right)\frac{1}{T}\right].
\end{eqnarray}

That means that at $350^\circ$C, the equilibrium between our
impinging flux and the evaporation of Cd in the nanoparticle is
realized at 2\% Cd; the equilibrium temperature rises above
$400^\circ$C at 1\% Cd. If the Cd concentration $c$ stays around
these values, we can neglect its influence on the value of the
activation energy in Eq.~\ref{Eq pCd}. Actually the order of
magnitude of $c$ can be obtained if we assume that all the Cd
contained in the exponential tail of length $\tau$ describing the
interface (Eq.~\ref{Eq interface}) is due to the reservoir effect in
the gold nanoparticle. Distributing the Cd content of a cylinder of
CdTe of height $\tau=1.6$~nm and diameter $d_{NW}=$~7.5~nm, into a
gold sphere of diameter $d_{NP}=~$13~nm (diameters measured on the
TEM image) results in a concentration equal to 1.5\% (a similar
value, 2\%, is calculated for the NW studied by EDX in
Fig.~\ref{fig1}). This is well below the concentration allowing the
formation of Cd-Au compounds. Note that this is an average
concentration; it gives no information about the distribution:
uniform, or with an axial gradient, or a radial one, or even limited
to the surface.

The sublimation of CdTe in the same temperature range is
significant. In addition to several reports on the Cd and Te vapor
pressure above CdTe, the sublimation rate from the (001) surface was
measured by RHEED oscillations \cite{Arias, Waag, Tatarenko}
together with the growth rate. The sublimation rate is
\cite{Arnoult} (in (001) ML s$^{-1})$:
$V_{sub}=3.1\times10^{-13}~\exp(-21600/T)$, with an activation
energy equal to 1.86 eV. This is close to the activation energy of
vapor pressure measured above CdTe \cite{Yang},
$p_{Cd}~p_{Te}^{\frac{1}{2}}\sim\exp (-33200/T)$, hence
$p_{Cd}~\sim\exp (-22100/T)$ under congruent sublimation. Although
the sublimation rate is expected to depend on the orientation of the
surface (its anisotropy gives rise to the Wulff construction of the
shape of a sublimating crystal), we will take the (001) rate as a
first estimate of the sublimation from the NW sidewalls. It is worth
at this point to remind the interpretation of the growth rate, which
is smaller than the incident flux (incorporation less than unity):
the difference involves two contributions, the sublimation, and the
desorption of the adatoms which do not reach a nucleation center. In
the case of 2D growth, nucleation centers are steps or stable nuclei
(larger than the critical size). In the classical
Burton-Cabrera-Frank analysis, which was applied to CdTe
\cite{Peyla, Pimpinelli}, the flux of adatoms towards the steps is
proportional to $(F-V_{sub}) \lambda$, where $F$ is the incident
flux and $V_{sub}$ the desorption rate, and the effective diffusion
length $\lambda$ takes into account the true diffusion length $x_s$
(desorption) and the average distance between nucleation centers
$l_s$ (incorporation): $\lambda=\frac{x_s}{l_s}\tanh
\frac{l_s}{x_s}$. In the case of a NW sidewall, the gold
nanoparticle forms another trap, and if its efficiency is similar to
that of the steps, it receives the same adatom flux $(F-V_{sub})
\lambda$. As a result, the sublimation makes the flux smaller,
through the factor $(F-V_{sub})$ instead of $F$. $F$ itself if quite
small if the flux-NW angle is small, and the lateral growth makes
$\lambda$ smaller either by the propagation of steps formed at the
NW-substrate interface \cite{Filimonov} or by the formation of
critical nuclei on the sidewalls as in 2D growth \cite{Peyla,
Pimpinelli}.

The contribution to the NW growth - positive if the diffusion takes
place from the sidewall to the nanoparticle, or negative if from the
nanoparticle to the sidewalls - is thus

\begin{equation}\label{Eq VSW}
(V_{SW}^{in}-V_{SW}^{out})=(\frac{J_s \tan \alpha}{\pi}-V_{sub})~ 4
\lambda / D_{NW}.
\end{equation}
With the present growth conditions ($J_s$=0.5 ML s$^{-1}$ and $\tan
\alpha$=0.21 ), the impinging flux is fully compensated by the
sublimation around $350^\circ$C.

Finally, we should calculate the concentration $c$ in the
nanoparticle by writing the equilibrium between:
\begin{itemize}
\item the sum of the four previous contributions,
$V_{NP}^{in}-V_{NP}^{out}+V_{SW}^{in}-V_{SW}^{out}$, where the first
two terms result from the direct flux to the nanoparticle and the
evaporation from the nanoparticle, and the last two terms are the
contribution of the diffusion between the sidewalls and the
nanoparticle.
\item the nucleation at the NW tip. In the case of the self-catalyzed growth of GaAs
NWs, the knowledge about the NW-nanodroplet interface is good
enough, that classical nucleation theory can be fully developed and
applied to calculate the nucleation rate as a function of the
droplet composition \cite{Glas2013}. It features a fast increase as
a function of the difference of chemical potential - \emph{i.e}, as
a function of the logarithm of the activity of As in the Ga droplet.
We will use a crude approximation, by assuming that the nucleation
rate is zero below a threshold $c_0$ (which reflects the equilibrium
concentration and the effect of the nucleation barrier) and then
rises linearly with $c$. We ignore the stochastic nature of the
nucleation \cite{Glas2017} since we calculate an average growth
rate.
\end{itemize}

\begin{figure}
 \centering
 \includegraphics[width=1\columnwidth]{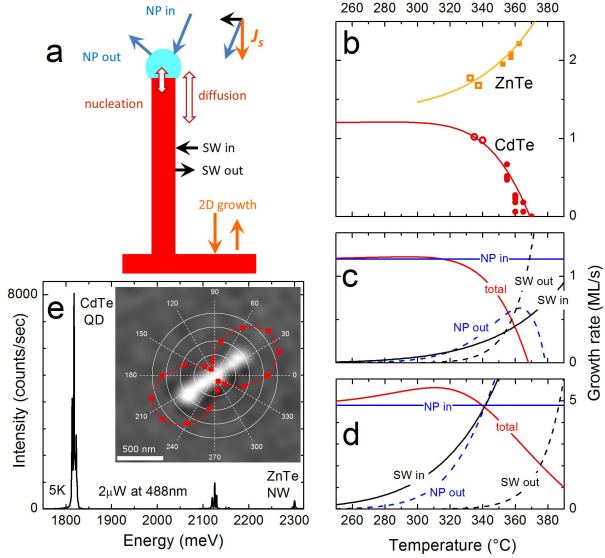}
\caption{(a) Scheme of the different contributions to the growth of
the NW; the role of the different components of the incident flux
(top right) is identified by their color (orange: normal component
$J_s$ contributing to the 2D growth and calibrated through RHEED
oscillations; black: component $J_s \tan \alpha$ normal to the NW
axis impinging the NW sidewalls; blue: total flux of intensity $J_s/
\cos \alpha$) (b) growth rate of the ZnTe and CdTe segments of
sample T (solid symbols) and another sample (open symbols), as a
function of temperature. The normal incident flux measured on test
(001) 2D layers is the same for both, 0.5 ML s$^{-1}$, with $\tan
\alpha =0.21$ for CdTe and 0.48 for ZnTe; the same unit is used for
the growth rate of CdTe and ZnTe in the NW; (c) the four components
entering the CdTe growth rate as calculated for the conditions in
(b); (d) the four components for a higher flux (2 ML s$^{-1}$) and
an oblique incidence ($\tan \alpha =1$); (e) photoluminescence
spectrum and (inset) polarization diagram superimposed on the image
of the NW.} \label{fig4}
\end{figure}

The result is shown in Fig.~\ref{fig4}b, where we plot the growth
rate of ZnTe and CdTe segments in sample T and a similar sample
grown at lower temperatures, as a function of temperature. For ZnTe
we show only the segments at high growth rate, which are associated
with quasi-full-sphere nanoparticles \cite{Rueda growth}, and the
fitting parameter is the diffusion length along the NW sidewalls. A
good fit is obtained by assuming an activation energy 0.95~eV. A
slow increase of the diffusion length with temperature was already
noted \cite{Rueda growth} in association with a smaller tapering.
For CdTe we keep the same geometrical parameters and diffusion
length value, correct for the different angle of incidence, and we
add the sublimation of CdTe and the evaporation of Cd from the gold
nanoparticle, taken from the literature as discussed previously.

\section{\label{Discussion}Discussion}

\subsection{\label{Rates}Growth rates}

The four contributions to the growth rate are detailed in
Fig.~\ref{fig4}c. At low temperature, the only significant
contribution is the direct flux to the nanoparticle. As the
temperature increases, the diffusion length also increases and so
does the contribution from the flux to the sidewall. However, this
positive flux is rapidly compensated by the evaporation from the
nanoparticle. In this temperature range, the flux is positive from
the NW sidewalls to the nanoparticle. At higher temperature, the
sublimation from the NW sidewalls becomes significant, so that the
adatom density decreases (blocking the formation of lateral dots)
and finally the flux is reversed, the Cd concentration in the
nanoparticle goes down, and the axial growth is stopped.  It appears
that the evaporation of Cd from the gold nanoparticle, the small
angle of incidence on the sidewalls and the sublimation of CdTe from
the sidewalls conspire to achieve a low growth rate, a low
concentration of Cd in the nanoparticle (resulting in sharp
interfaces, much sharper than reported in earlier studies
\cite{Dluzewski} realized in the VLS mode with Au-Ga liquid droplets
at $450^\circ$C), and a low adatom density on the sidewalls
(resulting in a weak lateral growth, as confirmed in the next
subsection).

Other reports on the growth of CdTe NWs by MBE
\cite{Wojnar2017,Kessel} mention higher values of flux (larger
values of the beam equivalent pressure on each of a Cd and a Te
cells, than that from a single CdTe cell in our case). Additionally,
the angle of incidence to the NW may be \cite{Wojnar2017}, or is on
purpose\cite{Kessel}, larger than ours. The temperature dependance
of each of the four contributions, calculated using the present
model and for such conditions, is shown in Fig.~\ref{fig4}d. The
result is a higher growth rate at $350^\circ$C and a similar growth
rate at $390^\circ$C. It is worth noting that the impact of the
sublimation from the sidewalls is significantly decreased (it is now
competing with a lateral flux with a higher intensity at relative
incidence closer to the normal) so that the adatom density is
expected to be larger than under our present growth conditions: in
Fig.~\ref{fig1} each CdTe insertion contributes by less than one
monolayer. This makes a sharp contrast with the thick shell
(equivalent to 2 nm of pure CdTe) measured by EDX \cite{Rueda2016}
on a sample we have grown previously using a high flux from a Cd
cell with a large angle of incidence. Such a strong lateral growth
is likely to give rise to the formation of additional quantum dots
which complicate the photoluminescence spectra and indeed have been
identified by cathodoluminescence \cite{Wojnar2016}.

\subsection{\label{PL}Luminescence}

As an example, Fig.~\ref{fig4}e shows the photoluminescence spectrum
of a CdTe-QD in ZnTe NW. The initial NW was grown at $375^\circ$C to
reduce tapering, then the CdTe insertion was grown at $340^\circ$C
as a compromise between a reasonably fast axial growth and a
reasonably slow radial growth. The whole structure was capped with a
(Zn,Mg)Te shell. The photoluminescence spectrum features three well
separated bands which we associate (for increasing energies) to the
long CdTe dot resulting from axial growth, to a parasitic CdTe
inclusion resulting from lateral growth, and to ZnTe. Note that the
polarization diagram shown in the inset is consistent with a
light-hole exciton  \cite{Jeannin}, while the polarization of the
intermediate line (not shown) is oblique, suggesting that the
principal axis of the insertion is not along the NW axis, and the
polarization of ZnTe (not shown) is consistent with a heavy-hole
exciton. Cathodoluminescence on the same NW (not shown) confirms
that the three lines originate from well-separated areas along the
NW.

\subsection{\label{Kirkendall}Negative growth and nanoparticle shape}

\begin{figure}
 \centering
 \includegraphics[width=0.9\columnwidth]{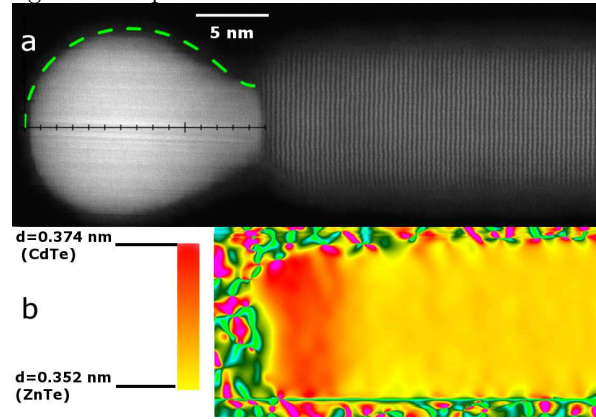} \caption{(a) TEM image of the tip of a NW
with the wurtzite structure, with a CdTe final segment as identified
by the GPA (b). The green dashed line shows the shape calculated
from the pendant droplet equation with $\beta=0.037$. The color code
is yellow for ZnTe (interplane distance 0.352 nm) and red for CdTe
(interplane distance 0.374 nm).} \label{fig5}
\end{figure}

The unstable character of the CdTe segments at such temperatures
(above $350^\circ$C) is also confirmed by the smaller length we
systematically observed on the final, uncapped CdTe insertions in
samples R and T. The CdTe segment is shorter while keeping the same
diameter, an effect similar to what has been reported as "negative
growth" upon annealing of GaAs NWs \cite{Dubrovskii2009}. In the
present case, it indicates that CdTe is re-dissolved into the gold
nanoparticle as soon as the impinging flux has been stopped,
creating a gradient of concentration, $\overrightarrow{\nabla}c$,
directed on average from the interface towards the apex of the
nanoparticle. This gradient drives the diffusion of Cd through the
nanoparticle, with a flux $-D_{Cd}\frac{\nabla c}{\Omega}$ where
$D_{Cd}$ is the diffusion coefficient of Cd in Au, and $\Omega$ the
atomic volume in gold ($=\frac{a_0^3}{4}$). The same gradient
induces self-diffusion of gold in the opposite direction, with a
flux $D_{Au}\frac{\nabla c}{\Omega}$. Diffusion in a substitutional
solid solution is accompanied by the so-called vacancy drift, which
is not balanced if the diffusion coefficients of the two atomic
species are different. In the present case, they are indeed very
different ($D_{Cd}/D_{Au}\approx8$ at 1000K) \cite{Decroupet} so
that a number of vacancies approximately equal to the number of
atoms dissolved is driven towards the interface, where they have to
diffuse to the surface. As a result the nanoparticle tends to assume
a bulb-like shape: this is indeed what we observe in Fig.~\ref{fig5}
for a NW with the wurtzite structure. This is a manifestation of the
Kirkendall effect, which has been known for years to create voids in
alloys, and has been used recently to fabricate hollow
nanostructures \cite{Gusak, Gonzalez}. Elongated shapes have been
observed during redissolution of GaAs, InAs or SnO$_2$ NWs into gold
nanoparticles upon heating \cite{Persson,Pennington, Hudak}. We
occasionally observe elongated nanoparticles, but the shape in
Fig.~\ref{fig5} is more complex and well approximated by the shape
of a pendant droplet \cite{Gassin,Rotenberg, Hansen}, with a value
of the characteristic parameter $\beta$=0.037. This is the signature
of a quasi-uniform force field, directed along the axis, induced
within the nanoparticle by the surface energy: in the pendant
droplet \cite{Gassin,Rotenberg,Hansen}, the force per unit volume is
$\beta\frac{\gamma}{R^2}$ where $R$ is the radius of curvature at
the apex, $\gamma$ is the energy per unit surface area; this force
compensates the difference of weight per unit volume between the
liquid in the droplet and the gas outside. The force corresponds to
the gradient of the Laplace pressure which is uniform in a spherical
droplet. In the context of nanowire growth, the equivalent of the
Laplace pressure is the shift of the chemical potential within the
catalyst droplet or nanoparticle (Gibbs-Thomson effect)
\cite{Glas2013}: in a spherical droplet, the shift is uniform, equal
to $\frac{2\gamma}{R} \Omega$. The surface energy along the
pendant-droplet shape results in a gradient of this chemical
potential, equal to $\beta\frac{\gamma}{R^2} \Omega$, which
decreases the Cd flux and increases the Au flux. The equilibrium
shape is achieved when the vacancy drift vanishes, \emph{i.e.}, when
the Cd (or Cd and Te) and Au currents compensate each other. Within
the regular solution model, the gradients of chemical potentials
become $\nabla\mu_{Cd}=k_BT\frac{\nabla
c}{c}-\beta\frac{\gamma}{R^2} \Omega$ and
$\nabla\mu_{Au}=-k_BT\frac{\nabla c}{1-c}+\beta\frac{\gamma}{R^2}
\Omega$, respectively. Balancing the diffusion flux of Cd and Au
implies $D_{Cd} \nabla\mu_{Cd}=-D_{Au} \nabla\mu_{Au}$, hence
$R\nabla c =\frac{D_{Cd}c+D_{Au}(1-c)}{D_{Cd}-D_{Au}}\beta
\frac{1}{k_BT}\frac{\gamma  \Omega}{R}$. With $c$ a few $\%$,
$D_{Cd}/D_{Au}\approx8$, $  \Omega=1.7\times10^{-2}$ nm$^3$,
$\gamma$=1.5 J m$^{-2}$ for gold \cite{Vitos}, and $k_BT$=50 meV, we
obtain $R\nabla c\approx10^{-3}$: the concentration drop between the
interface and the apex is a (significant) fraction of the
concentration estimated previously. This estimate is certainly
oversimplified (we totally ignore the role of Te, and the actual
flux distributions are more complex) but it shows that the orders of
magnitude are realistic.

\section{\label{Summary}Summary}

CdTe QDs have been grown in ZnTe NWs, with abrupt interfaces and a
controlled aspect ratio, and reduced parasitic growth on the
sidewalls. This is achieved in VSS growth by MBE using a gold
catalyst and a low flux of Cd and Te, quasi-parallel to the NW axis,
at temperatures where a simple model shows that the sublimation of
CdTe and the evaporation of Cd from the gold nanoparticle are
significant. Similar concepts should apply to the growth of other
systems, for instance NWs incorporating CdSe or HgTe. Very peculiar
shapes of the gold nanoparticle are ascribed to the redissolution of
CdTe under these specific conditions which - through the Kirkendall
effect - favors the build up of a gradient of the chemical potential
along the nanoparticle.

\begin{acknowledgments}

This work was performed in the joint CNRS-CEA group "Nanophysique \&
semiconducteurs", the team "Laboratory of Material Study by Advanced
Microscopy", and the team "Materials, Radiations, Structure". We
benefitted from the access to the Nano-characterization facility
(PFNC) at CEA Minatec Grenoble. We acknowledge funding by the French
National Research Agency (projects Magwires, ANR-11-BS10-013,
COSMOS, ANR-12-JS10-0002, Espadon, ANR-15-CE24-0029 and labex LANEF
ANR-10-LABX-51-01).

\end{acknowledgments}

\end{document}